\documentclass[sigconf]{acmart}
\usepackage{soul} % 下划线命令改为\ul{}，解决换行问题
\usepackage{graphicx}
\usepackage{booktabs}
\usepackage{float}
\usepackage{overpic}
\usepackage{float}  %设置图片浮动位置的宏包
\usepackage{subfloat}
\usepackage{stfloats} % for pic
\usepackage[skip=0.5\baselineskip]{caption}
\usepackage{bm}
\usepackage{amsmath}
\usepackage{booktabs}
\usepackage{multirow}
\usepackage{multicol}
\usepackage{enumitem}
\usepackage{subcaption}
\usepackage{threeparttable}
\usepackage{xspace}
\usepackage{color}
\usepackage[normalem]{ulem}
\usepackage{algorithmicx,algorithm}
\usepackage{algpseudocode}
\usepackage{algorithm,algpseudocode}
\usepackage{marvosym}   % 添加通讯作者的信封
\usepackage{colortbl}   % 表格中的阴影颜色
\usepackage{tcolorbox}  % 文本加框
\tcbuselibrary{breakable}   % 配合文本加框，使得框可以跨页
\usepackage{wasysym}    % 添加实心圆空心圆
\usepackage[figuresright]{rotating} % 旋转表格
\usepackage{pifont} % 圆圈序号
\usepackage{tcolorbox}

\usepackage{caption} % 如果已加载则忽略
\title{DMESR: Dual-view MLLM-based Enhancing Framework for Multimodal Sequential Recommendation}
\author{Mingyao Huang$^{1}$\dag,Qidong Liu$^{1}$\dag, Wenxuan Yang$^1$, Moranxin Wang$^1$, Yuqi Sun$^1$, \\ 
Haiping Zhu$^{1}$, Feng Tian$^{1}$, Yan Chen$^{1}$}
\affiliation{
    \institution{$^1$ Schoole of Computer Science and Technology, Xi'an Jiaotong University}
    \country{}
}
\email{{liuqidong,zhuhaiping}@xjtu.edu.cn, {2214312088,yangwenxuan,wangmo,YuqiSun}@stu.xjtu.edu.cn}
\email{{fengtian,chenyan}@mail.xjtu.edu.cn}
\thanks{\dag~ Both authors contributed to this paper equally.}
\renewcommand{\shortauthors}{Mingyao and Qidong et al.}

\begin{document}
\begin{sloppypar} 

\begin{abstract}
    Sequential Recommender Systems (SRS) aim to predict users' next interaction based on their historical behaviors,
    while still facing the challenge of data sparsity.
    %and have become one of the cornerstones in modern recommender systems. 
    With the rapid advancement of Multimodal Large Language Models (MLLMs), leveraging their multimodal understanding capabilities to enrich item semantic representation has emerged as an effective enhancement strategy for SRS. However, existing MLLM-enhanced recommendation methods still suffer from two key limitations. First, they struggle to effectively align multimodal representations, leading to suboptimal utilization of semantic information across modalities. Second, they often overly rely on MLLM-generated content while overlooking the fine-grained semantic cues contained in the original textual data of items. To address these issues, we propose a \textbf{D}ual-view \textbf{M}LLM-based \textbf{E}nhancing framework for multimodal \textbf{S}equential \textbf{R}ecommendation\ (\textbf{DMESR}). 
    %MLLM-DVIR employs a contrastive learning mechanism to semantically align multimodal text representations generated by MLLMs, and introduces a cross-attention fusion module to integrate the coarse-grained semantic knowledge refined by MLLMs with the fine-grained original textual semantics.
    For the misalignment issue, we employ a contrastive learning mechanism to align the cross-modal semantic representations generated by MLLMs.
    For the loss of fine-grained semantics, we introduce a cross-attention fusion module that integrates the coarse-grained semantic knowledge obtained from MLLMs with the fine-grained original textual semantics.
    Finally, these two fused representations can be seamlessly integrated into the downstream sequential recommendation models. 
    Extensive experiments conducted on three real-world datasets and three popular sequential recommendation architectures demonstrate the superior effectiveness and generalizability of our proposed approach.
    % To ease the reproducibility, we have released the code online\footnote{https://github.com/mingyao-huang/DMESR.git}.
\end{abstract}

%% CCS Concepts



\maketitle

\section{Introduction} \label{sec:introduction}

Sequential recommender systems have emerged as an important branch of recommender systems, aiming to predict the next item a user is most likely to interact with based on their historical interaction sequence~\cite{fang2020deep,wang2019sequential}. With the rapid development of deep learning techniques, SRS has been successfully applied in various domains such as e-commerce~\cite{wang2020time} and short-video platforms~\cite{pan2023understanding}. The core challenge of SRS lies in effectively modeling users' dynamic and evolving preferences from their interaction sequence.
Benefiting from the progress of deep learning, many recent SRS models, e.g., SASRec~\cite{sun2019bert4rec} and BERT4Rec~\cite{kang2018self}, have demonstrated strong capability in recommendation. 
However, they primarily rely on interaction sequences, limiting their ability to capture rich item semantics.
%However, they still suffer from data sparsity and cold-start problems~\cite{zhou2016evaluating}.
%and online education~\cite{zhang2019hierarchical}

To alleviate these issues, multimodal information, e.g., text, images, and videos, was introduced into sequential recommendation~\cite{he2016vbpr,hu2023multi}. %This line of work provided richer contextual cues for understanding user preferences and complemented traditional ID- or text-only representations. 
Incorporating such multimodal signals enables the learning of more informative item representations, thereby providing richer contextual cues for understanding user preferences.
Currently, many multimodal recommender systems (MRS) have been proposed to leverage these heterogeneous modalities through joint embedding~\cite{he2016vbpr}, knowledge graph~\cite{sun2020multi}, or attention fusion~\cite{liu2022disentangled}. %Despite their effectiveness, these methods often struggle to deeply integrate various modalities due to their limited semantic reasoning capability and reliance on modality-specific encoders, which may lead to \lqd{\textbf{Insufficient Fusion}} and \textbf{Semantic Inconsistency}~\cite{zhou2023comprehensive,xu2025survey}.
Despite their effectiveness, these methods often struggle to deeply integrate various modalities. Their limited semantic understanding capability makes it difficult to capture high-level cross-modal relations, leading to \textbf{Insufficient Semantic Exploitation}. Besides, their reliance on modality-specific encoders causes the representations of different modalities to be distributed in heterogeneous semantic spaces, resulting in \textbf{Semantic Inconsistency}~\cite{zhou2023comprehensive,xu2025survey}.
With the rapid advancement of Multimodal Large Language Models (MLLMs), such as GPT-4V~\cite{yang2023dawn} and Qwen2.5-VL~\cite{bai2025qwen2}, these models have demonstrated remarkable capability in merging multimodal information into a unified textual semantic space. 
Thus, owing to their strong ability to comprehend and fuse heterogeneous modalities, MLLMs are promising to alleviate the issues of insufficient semantic exploitation and inconsistency, showing great potential for enhancing SRS~\cite{liu2024rec,wang2025leveraging}.
%By bridging textual and visual modalities, MLLMs greatly enhance the system’s ability to comprehend and reason over complex data inputs, thereby significantly improving the accuracy of downstream recommendation tasks~\cite{liu2024rec,wang2025leveraging}. 
% Building on this strength, their multimodal semantic understanding has been increasingly adopted to enrich item representations, which has proven an effective approach to improving sequential recommendation performance~\cite{pomo2025recommender,ye2025multimodal,liu2024rec}.

Recent studies have investigated the integration of MLLMs into recommender systems from different perspectives. Some works~\cite{liu2024rec,ye2025harnessing,tourani2025rag,zhang2025histllm} directly employ MLLMs as the recommender backbone, where item multimodal data are first processed by the MLLM into textual or tokenized representations, and the user's interaction history is then fed into the large model to infer user preferences and rank candidate items. 
Although these methods exhibit strong understanding ability, they suffer from extremely high computational costs during training and inference process, making them impractical for real-time industrial applications~\cite{geng2024breaking}. 
To address this, another line of research studies~\cite{lyu2024x,mo2025one,luo2024qarm,luo2024molar,tian2024mmrec} focuses on incorporating MLLM-derived semantics into downstream recommendation models, where pre-trained MLLMs are only used during training. 
% to generate multimodal item representations that are cached for later usage. 
% During inference, the recommendation model predicts user–item interactions without invoking the MLLM, thereby meeting the real-time requirements of recommendation services. 
Despite their success, current methods still face two critical issues:
%\textbf{(i) Unaligned Multimodal Semantic Spaces.} 
\textbf{(i) Inconsistent Cross-modal Semantics}. Many approaches~\cite{lyu2024x,tian2024mmrec,zhang2025histllm,liu2024harnessing} simply concatenate or average MLLM outputs of different modalities, neglecting the intrinsic misalignment between textual and visual semantic spaces. This leads to suboptimal fusion and limited recommendation gain.
\textbf{(ii) Loss of Fine-grained Semantics.} Most methods over-rely on MLLM-generated content obtained by paraphrasing the original item texts~\cite{lyu2024x,ye2025harnessing,luo2024molar}. However, the paraphrasing process may result in loss of fine-grained and domain-specific details, leaving the original data underutilized. 

To address these challenges, we propose a \textbf{D}ual-view \textbf{M}LLM-based \textbf{E}nhancing framework for multimodal \textbf{S}equential \textbf{R}ecommen\-dation\ (\textbf{DMESR}).
% The dual-view enhancing framework consists of the \textit{Coarse-Grained Semantic View} and the \textit{Fine-Grained Semantic View}. For the coarse-grained view, we design a three-way prompting framework that guides MLLMs to capture abstract, coarse-grained semantics from multimodal item data. To mitigate the issue of inconsistent cross-modal semantics, we employ a contrastive learning module to align the modality-specific embeddings into a unified semantic space. 
% For the fine-grained view, we directly encode the raw textual descriptions of items to preserve detailed and domain-specific semantics.
% \lqd{To effectively combine these two views}, we design a Bidirectional Cross-Attention Fusion Module, which enables mutual enhancement between the two semantic views. 
The dual-view enhancing framework consists of the \textit{Coarse-Grained Semantic View} and the \textit{Fine-Grained Semantic View}.  
For the coarse-grained view, we propose a \textbf{Cross-modal Semantic Derivation} stage, where a three-way prompting framework is designed to guide MLLMs to capture abstract semantics from multimodal item data. Then, to mitigate the issue of inconsistent cross-modal semantics, a contrastive learning module is further employed to align the modality-specific embeddings.
For the fine-grained view, we introduce a \textbf{Bidirectional Semantic Fusion} stage, where the raw textual imformation of items are directly encoded to preserve detailed semantics. We then apply a bidirectional cross-attention fusion module to integrate the fine-grained textual semantics with the coarse-grained MLLM semantics between these two views.  
Finally, both fused semantic representations are fed into downstream sequential recommendation models.
The major contributions of this paper are summarized as follows:
\begin{itemize}[leftmargin=*, topsep=4pt, after=\vspace{-4pt},]
\item We propose a dual-view MLLM-based enhancing framework, which jointly models coarse- and fine-grained item semantics for multimodal sequential recommendation.
\item We design a contrastive alignment module to align cross-modal representations and a bidirectional cross-attention module to fuse dual-view semantics.
\item We conduct extensive experiments on three real-world datasets with three backbone SRS models to validate the effectiveness and flexibility of DMESR.
\end{itemize}

\section{Preliminary} \label{sec:preliminary}

% \textbf{Sequential Recommendation.}
The sequential recommendation task aims to predict the next item that a user is most likely to interact with based on their historical interaction sequence. 
Let $\mathcal{U} = \{u_1, u_2, ..., u_{|\mathcal{U}|}\}$ and $\mathcal{V} = \{v_1, v_2, ..., v_{|\mathcal{V}|}\}$ denote the sets of users and items, respectively. 
For each user $u \in \mathcal{U}$, the interaction sequence is represented as $S_u = \{v^{(u)}_1, v^{(u)}_2, ..., v^{(u)}_{n_u}\}$, ordered chronologically, where $n_u$ is the length of the user’s historical interactions.
For simplicity, we omit the superscript $(u)$ in the following parts of this paper.
The goal of sequential recommendation is to predict the most probable next item $v_{n_u+1}$ that the user will engage with, which can be formulated as: $v_{n_u+1} = \arg\max_{v_i \in \mathcal{V}} P(v_{n_u+1} = v_i \mid S_u)$.

\section{Method} \label{sec:method}

\subsection{Overview}

\begin{figure*}[t]
    \centering
    \includegraphics[width=0.85\textwidth]{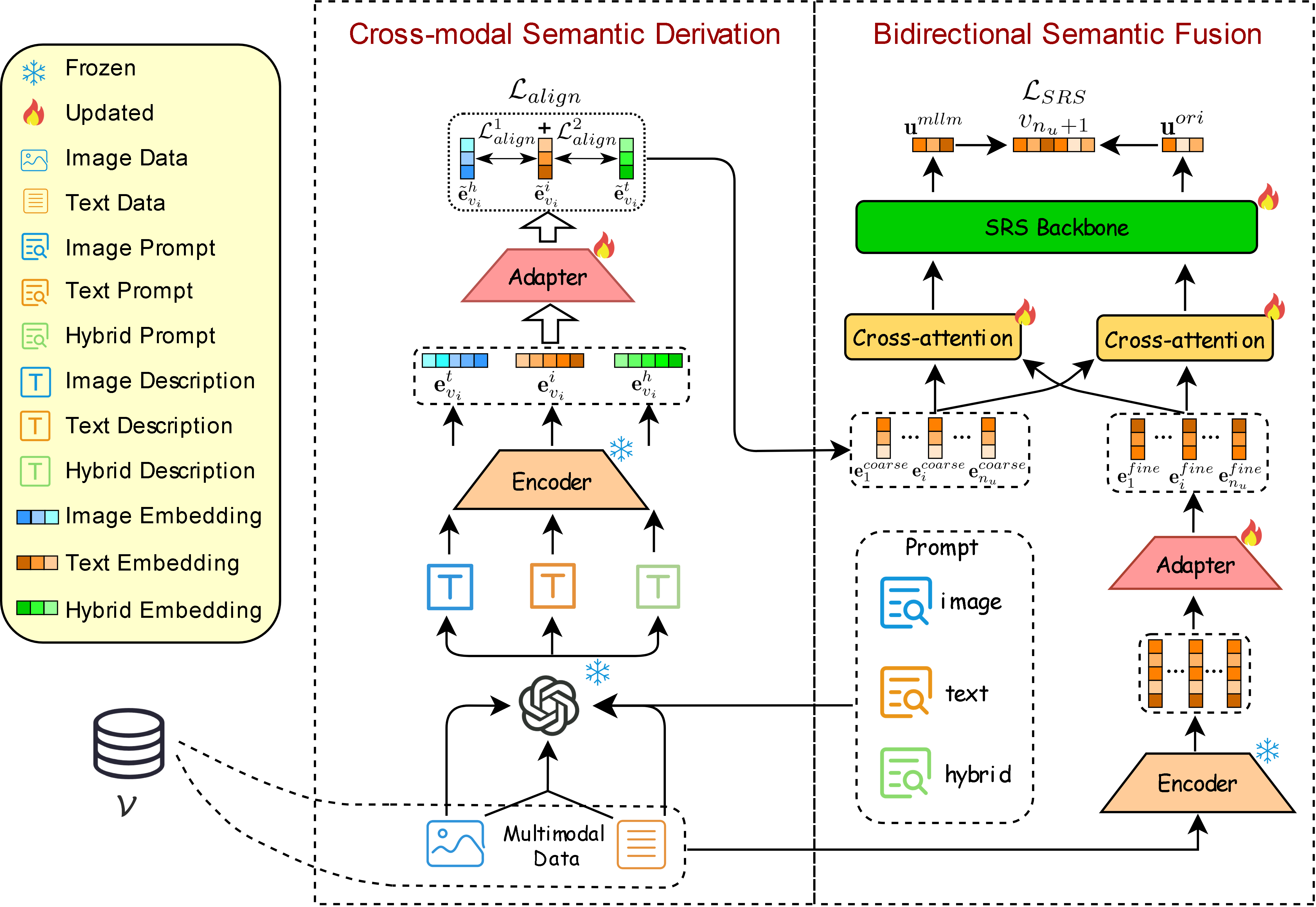}
    \caption{The overall architecture of the proposed DMESR framework. 
    %It consists of two major stages: (1) Cross-modal Semantic Derivation and (2) Bidirectional Semantic Fusion.
    }
    \label{fig:framework}
    \vspace{-2mm}
\end{figure*}

Figure~\ref{fig:framework} illustrates the overall workflow of the proposed \textbf{DMESR} framework, which enhances sequential recommendation by integrating coarse-grained multimodal semantics from MLLMs with fine-grained semantics from original texts. The framework operates in two stages. 
In the first stage, known as %\textbf{Multimodal Semantic Representation Learning}
%\hmy{\textbf{Multimodal Semantic Alignment}}, 
\textbf{Cross-modal Semantic Derivation}. 
The objective is to leverage the strong multimodal understanding capability of MLLMs to construct rich item coarse-grained representations. Specifically, a three-way prompting framework is designed to guide the MLLM to generate semantic descriptions from different modalities. 
To ensure consistency among different modalities, a contrastive learning module is further designed to align cross-modal embeddings within a unified semantic space.
The second stage, termed
%\textbf{Cross Semantic Fusion}, 
\textbf{Bidirectional Semantic Fusion},
aims to complement the coarse-grained MLLM semantics with the fine-grained and domain-specific semantics preserved in the original item texts. To achieve this, the raw textual data are directly encoded into embeddings. Then, a bidirectional cross-attention module is introduced to enable semantic interaction between these two views of embeddings. During the inference phase, DMESR generates all these item embeddings in advance. Then, these precomputed embeddings replace the original embedding layer of the SRS to achieve effective utilization.
%, ensuring both inference efficiency and semantic enhancement.

% -------------------------------
\subsection{Cross-modal Semantic Derivation}

To obtain aligned multimodal semantic representations, this stage first introduces a \textbf{Three-way Prompting Framework} to generate multimodal textual descriptions from the MLLM.
Then, it employs an \textbf{Adapter Network} to inject collaborative signals while reducing embedding dimensionality.
Finally, a \textbf{Cross-modal Representation Alignment} module is proposed to unify the three types of embeddings into a shared semantic space. Through this stage, we derive a unified coarse-grained multimodal item representation.
% -------------------------------

% \vspace{2mm}
\subsubsection{\textbf{Three-way Prompting Framework}}
% \subsubsection{Three-way Prompting Framework.}
To fully utilize the multimodal reasoning ability of MLLMs, we construct a three-way prompting framework consisting of textual, visual, and hybrid (text+image) prompting routes for MLLM Semantic Generation. The three prompting routes are defined as follows:

\vspace{1mm}
\noindent\textit{\textbf{Textual Prompting.}}  
The textual prompting route leverages the MLLM’s semantic understanding in pure text scenarios to enrich the original textual descriptions of items. Each item $v$ is represented by a set of textual attribute pairs $\{(a_1, v_1), (a_2, v_2), ..., (a_n, v_n)\}$.  
Accordingly, the textual prompt explicitly enumerates these attribute pairs and instructs the MLLM to generate a concise description for recommendation. %Due to the space limitation, we place all detailed prompt templates in the \textbf{Code Repository}.
The text prompt $T_{v_i}$ is designed as follows:
%\lqd{$\{\mathrm{MLLM}(\text{Prompt}_{{text}}(T_{v_i}))\}_{v_i \in \mathcal{V}} = D_{{text}}$}
\begin{tcolorbox}[colframe=gray!50!black, colback=white, coltitle=white, title=Textual Prompt,  boxsep=1pt,
    top=0pt, bottom=0pt,]\textit{The item has the following attributes: $attribute_1: <content_1>; attribute_2: <content_2>; ...; attribute_n: <content_n>$. Summarize the item based on the text above and describe it smoothly in one paragraph for recommendation.}\end{tcolorbox}
\noindent Denote the textual prompting process as   
$\mathcal{D}_{\text{text}} = \{\mathrm{MLLM}(\text{Prompt}_{text}(T_{v_i})) \mid v_i \in \mathcal{V}\}$, where $\mathrm{MLLM}(\cdot)$ represents the reasoning process of the MLLM, and $\mathcal{D}_{{text}}$ denotes the set of all generated textual descriptions.  
Each generated description $d_{v_i}^t \in \mathcal{D}_{text}$ is then encoded as $\mathbf{e}_{v_i}^t = \mathrm{Emb}(d_{v_i}^t)$, where $\mathrm{Emb}(\cdot)$ denotes an LLM-based encoder implemented by \texttt{text-embedding-ada-002}, and $\mathbf{e}_{v_i}^t \in \mathbb{R}^{d_0}$ represents the resulting semantic embedding, with $d_0$ being the output dimensionality.
%\lqd{$\mathbf{e}$}

\vspace{1mm}
\noindent\textit{\textbf{Visual Prompting.}}  
The visual prompting route utilizes only the item's image modality to explore visual information that benefits recommendation.  
To guide the MLLM in mining visual cues, the visual prompt explicitly directs the model to focus on elements such as color schemes and main characters to generate a vivid descriptive paragraph.
Specifically, the visual prompt template is as follows:
\begin{tcolorbox}[colframe=gray!50!black, colback=white, coltitle=white, title=Visual Prompt, top=2pt, bottom=2pt, boxsep=1pt, breakable]{\textit{Image data is given via local path \{image path\}. Analyze the provided image carefully, pay close attention to specific visual elements such as color schemes, main characters, prominent objects, and artistic style, and explicitly mention them in your description. The description should accurately reflect the image's theme, mood and atmosphere. Output only a well-structured and vivid paragraph for recommandation.}}\end{tcolorbox}
\noindent The placeholder \{image path\} will be replaced by the actual image file path. 
Denote the visual prompting process as   
$\mathcal{D}_{image} = \{\mathrm{MLLM}(\text{Prompt}_{image}(I_{v_i})) \mid v_i \in \mathcal{V}\}$,  
where $I_{v_i}$ denotes the image data path of item $v_i$, and $\mathcal{D}_{{image}}$ denotes the corresponding set of image-based descriptions.  
Similar to the textual promopting, each generated description $d_{v_i}^i \in D_{image}$ is encoded as $\mathbf{e}_{v_i}^i = \mathrm{Emb}(d_{v_i}^i)$.

\vspace{1mm}
\noindent\textit{\textbf{Hybrid Prompting.}}  
While the above two routes capture single-modality semantics, they overlook the complementary information between text and image, leading to incomplete multimodal understanding.  
To address this limitation, we design a hybrid prompting route that simultaneously inputs both textual and visual information into the MLLM.  
To fully exploit the multimodal understanding capability of the MLLM, we further devise a Recursive Prompting (RP) strategy ~\cite{dua2022successive} for prompt design.
This strategy iteratively refines the generated description through the following three steps:
\begin{itemize}[leftmargin=*,topsep=2pt,itemsep=1pt,]
    \item \textbf{Initial Generation.} Prompt the MLLM to jointly analyze the item’s text and image, focusing on correlations between textual attributes and visual cues, and generate an initial description.
    \item \textbf{Recursive Refinement.} Reanalyze the original modalities and the current description to detect missing, inconsistent, or ambiguous content. Revise the description accordingly to improve completeness and consistency.
    \item \textbf{Final Output.} Repeat Step 2 until no new information emerges, yielding a coherent and high-quality multimodal description.
\end{itemize}

% Guided by the design above, the Hybrid prompt is constructed as follows:

\begin{tcolorbox}[colframe=gray!50!black, colback=white, coltitle=white, title=Hybrid Prompt, top=2pt, bottom=2pt, boxsep=1pt, breakable]{
The item's text info is: \{text info\}. The image is given in \{image path\}.
Carefully analyze the provided textual description and movie poster through the following iterative steps.

\textbf{Step 1: Initial Generation.}
Jointly analyze the item's text and image information to identify key thematic and visual elements relevant to recommendation, such as genre, emotional tone, main characters, setting, and target audience.
Generate an initial paragraph that integrates the original textual content with complementary visual cues.

\textbf{Step 2: Recursive Refinement.}
Review the generated paragraph and reanalyze original modalities to detect missing or inconsistent details.
Revise the description by adding omitted but relevant visual or textual information to improve completeness, coherence, and recommendation focus.

\textbf{Step 3: Final Output.}
Repeat Step 2 until no new information added, yielding a unified, recommendation-oriented description.}
\end{tcolorbox}

%The complete prompt design of the RP strategy is provided in Appendix.
\noindent Denote the Hybrid prompting process as  
$D_{hybrid} = \{\mathrm{MLLM}(\text{Prompt}_{hybrid}(T_{v_i}, I_{v_i})) \mid v_i \in \mathcal{V}\}$, 
where $D_{hybrid}$ represents the set of final multimodal descriptions of all items.
Each description $d_{v_i}^h \in D_{hybrid} $ is then encoded as $\mathbf{e}_{v_i}^h = \mathrm{Emb}(d_{v_i}^h)$.

% Through this three-way prompting framework, semantic embeddings of items are obtained from three complementary perspectives, providing rich multimodal representations for subsequent recommendation.
% -------------------------------
%\textbf{\textit{Adapter.}}  
% \subsubsection{Adapter Network.}
% \vspace{2mm}
\subsubsection{\textbf{Adapter Network}}
Although the embeddings derived from three-way prompting framework contain rich modality-enhanced semantics, they lack the collaborative signals that are crucial for effective recommendation. Meanwhile, the high-dimensional output embeddings from the encoder are incompatible with the relatively low-dimensional embedding space of SRS models. 
Thus, we introduce lightweight adapter networks to project the semantic embeddings $\mathbf{e}^t$, $\mathbf{e}^i$ and $\mathbf{e}^h$ into a compact space while injecting collaborative signals during training. 
% Essentially, the adapter module is implemented as two linear fully connected layers. The first layer maps the input embedding from its original dimension to a half-sized latent space, and the second layer further projects it into the embedding space. 
For example, the adapted embedding for the text route $\mathbf{e}_{v_i}^t$ can be derived by:
\begin{equation}
\tilde{\mathbf{e}}_{v_i}^t = \mathbf{W}_2 \cdot (\mathbf{W}_1 \cdot \mathbf{e}_{v_i}^t + \mathbf{b}_1) + \mathbf{b}_2
\end{equation}
where $\mathbf{W}_1 \in \mathbb{R}^{d_1 \times d_0}$ and $\mathbf{W}_2 \in \mathbb{R}^{d \times d_1}$ are linear transformation matrices,  
$d_1 = \frac{1}{2} d_0$ is the intermediate hidden size,  
and $d$ denotes the target embedding dimension,   $\mathbf{b}_1 \in \mathbb{R}^{d_1}$ and $\mathbf{b}_2 \in \mathbb{R}^{{d}}$ are bias terms.  
%Denote this adaptation process as $Adapter_1^*(\cdot)$.
Similarly, the adapted embeddings for the visual $\mathbf{e}_{v_i}^i$ and hybrid routes $\mathbf{e}_{v_i}^h$ can be obtained by:
$\tilde{\mathbf{e}}_{v_i}^i = \text{Adapter}_1^i(\mathbf{e}_{v_i}^i)$, 
$\tilde{\mathbf{e}}_{v_i}^h = \text{Adapter}_1^h(\mathbf{e}_{v_i}^h)$.
The adapter parameters are jointly optimized with the main SRS model, which enables it to achieve dimensionality reduction while capturing collaborative dependencies.
%, thereby enhancing recommendation effectiveness.
% -------------------------------
% \subsubsection{Multimodal Representation Alignment.}

% \vspace{2mm}
\subsubsection{\textbf{Cross-modal Representation Alignment}}
While the adapter networks effectively project the multimodal embeddings into a compact and collaborative space, the representations derived from different prompting routes still reside in heterogeneous semantic spaces and exhibit semantic gaps, resulting in inconsistent multimodal representations.  
To address this issue, we introduce a contrastive alignment module that aligns the three embeddings into a unified semantic space.

For the three types of adapted item embeddings $\tilde{\mathbf{e}}_{v_i}^{t}$, $\tilde{\mathbf{e}}_{v_i}^{i}$, and $\tilde{\mathbf{e}}_{v_i}^{h}$ obtaining from the adapter network, we employ an in-batch contrastive learning to align them, where the objective is to maximize the similarity between embeddings of the same item while minimizing the similarity with other items within the batch.
For example, the in-batch contrastive loss between the textual embeddings $\tilde{\mathbf{e}}_{v_i}^{t}$ and visual embeddings $\tilde{\mathbf{e}}_{v_i}^{i}$ is formulated as:
\begin{equation}
\mathcal{L}_{{align}}^{1} = -\frac{1}{B} \sum_{i=1}^{B} \log 
\frac{\exp(\text{sim}(\tilde{\mathbf{e}}_{v_i}^{t}, \tilde{\mathbf{e}}_{v_i}^{i}))}
{\sum_{k=1}^{B} I_{[k \neq i]} \exp(\text{sim}(\tilde{\mathbf{e}}_{v_i}^{t}, \tilde{\mathbf{e}}_{v_k}^{i}))}
\end{equation}
where $B$ is the batch size, $\text{sim}(\cdot)$ denotes cosine similarity, and $I_{[k \neq i]}$ is 1 if $k \neq i$, otherwise 0.  
Swapping $\tilde{\mathbf{e}}_{v_i}^{t}$ and $\tilde{\mathbf{e}}_{v_i}^{i}$ in the above equation yields the other side of the contrastive loss $\mathcal{L}_{\text{align}}^{2}$, and the final contrastive loss between the textual and visual embeddings is defined as $\mathcal{L}_{\text{align}}^{(t \leftrightarrow i)} = \mathcal{L}_{\text{align}}^{1} + \mathcal{L}_{\text{align}}^{2}$.

Similarly, the alignment between the textual and hybrid embeddings is computed as $\mathcal{L}_{\text{align}}^{(t \leftrightarrow h)}$. Since textual data typically contains the most accurate and recommendation-oriented features (e.g., category, tags, and intro), we align the visual and hybrid modalities towards the textual space to form a unified semantic representation.  
Thus, the total multimodal alignment loss is expressed as:
\begin{equation}
\label{eq:contrastive_loss}
\mathcal{L}_{{align}} = \mathcal{L}_{{align}}^{(t \leftrightarrow i)} + \mathcal{L}_{{align}}^{(t \leftrightarrow h)}
\end{equation}

Through the contrastive alignment module, the textual embedding $\tilde{\mathbf{e}}_{v_i}^{t}$ serves as the semantic anchor and absorbs multimodal information from the visual and hybrid embeddings. Therefore, we directly adopt it as the unified coarse-grained multimodal representation, i.e., $\mathbf{e}_{v_i}^{coarse} = \tilde{\mathbf{e}}_{v_i}^{t}$, which is subsequently utilized in downstream recommendation tasks.
%for every item $v_i$, we obtain semantically unified multimodal item representations, denoted as $e_{v_i}^{mllm}$,

% -------------------------------
\subsection{Bidirectional Semantic Fusion}

Although MLLMs enrich item representations with contextual and world knowledge, the generated semantics may omit certain fine-grained details that are crucial for effective recommendation.  
For instance, an MLLM-generated description of a restaurant may capture its overall ambiance and cuisine style but overlook key details such as specific dish names or flavor features that strongly influence user preferences.  
%To compensate for this limitation, we design a \textbf{Bidirectional Cross-attention Fusion} module, which aims to integrate the MLLM-enhanced coarse-grained semantics with fine-grained original semantics. This is achieved by introducing the original textual information of items and applying a cross-attention module to enable mutual integration between the two semantic spaces, thereby achieving coarse–fine semantic fusion for collaborative recommendation.
To compensate for this limitation, this stage first performs a \textbf{Fine-grained Semantic Generation} process to extract detailed and domain-specific semantics from the original item texts.  
Then, a \textbf{Bidirectional Cross-attention Fusion} module is introduced to enable mutual integration between the coarse- and fine-grained semantic spaces.  
Finally, we conduct a \textbf{Collaborative Recommendation} to jointly model user preferences from both semantic views for final prediction.

% -------------------------------
% \subsubsection{Fine-grained Semantic Generation.}
% \vspace{2mm}
\subsubsection{\textbf{Fine-grained Semantic Generation}}
The fine-grained semantics are primarily contained in the original item texts, which are usually provided by merchants or product platforms. Such texts tend to be concise, accurate, and domain-specific, often including critical fields such as category, tags, and detailed descriptions that precisely characterize each item.  
% These properties allow the original text to capture fine-grained discriminative features that help distinguish items and enhance recommendation accuracy.
Similar to the process mentioned in the last section, %we encode the fine-grained original text $T_{v_i}$ as follows: %Formally, we encode the original text $T_{v_i}$ of item $v_i$ using the same textual encoder and then apply the adapter network for dimensional reduction and collaborative adaptation:
we first obtain the original textual embedding $\mathbf{e}_{v_i}^{T} = \text{Emb}(T_{v_i})$, and then apply the adapter network to produce the fine-grained representation:
\begin{equation}
\mathbf{e}_{v_i}^{fine} = \text{Adapter}_{2}(\mathbf{e}_{v_i}^{T}).
\end{equation}
Through this process, we obtain the fine-grained textual embeddings that preserve detailed semantic information of items, serving as semantic complements to the multimodal representations.

% -------------------------------
% \subsubsection{Bidirectional Cross-Attention Fusion.}
% \vspace{2mm}
\subsubsection{\textbf{Bidirectional Cross-Attention Fusion}}
After the previous process, we obtain the coarse-grained semantic embeddings from the view of MLLM and the fine-grained semantic embeddings from the view of the original texts. To achieve effective semantic fusion between these two views, we introduce a bidirectional cross-attention fusion module, which integrates these two semantic representations by performing cross-attention in both directions.

Taking the MLLM view as an example, for a user $u$ with the interaction sequence of $S_u=\{v_1,v_2,...,v_{n_u}\}$, we denote the two types of embeddings as
$\mathcal{E}_{u}^{coarse}=\{\mathbf{e}_{v_1}^{coarse}, \mathbf{e}_{v_2}^{coarse}, \ldots, \mathbf{e}_{v_{n_u}}^{coarse}\}$ and
$\mathcal{E}_u^{fine}=\{\mathbf{e}_{v_1}^{fine}, \mathbf{e}_{v_2}^{fine}, \ldots, \mathbf{e}_{v_{n_u}}^{fine}\}$.  
The coarse-grained-side refined embeddings are computed as:

\begin{equation}
\label{eq:cross_attn}
\begin{gathered}
Q = \mathcal{E}_{u}^{fine}\mathbf{W}^Q, \quad K = \mathcal{E}_{u}^{coarse}\mathbf{W}^K, \quad V = \mathcal{E}_{u}^{coarse}\mathbf{W}^V, \\
\hat{\mathcal{E}}_u^{coarse} = \text{Softmax}\!\left(\frac{QK^{\top}}{\sqrt{d}}\right)V
\end{gathered}
\end{equation}

\noindent where $\mathbf{W}^Q,\mathbf{W}^K,\mathbf{W}^V\in \mathbb{R}^{d\times d}$ are learnable parameters. By swapping $\mathcal{E}_u^{coarse}$ and $\mathcal{E}_u^{fine}$, we obtain the refined fine-gained embedding $\hat{E}_u^{fine}$ symmetrically.
%in a symmetric manner. 
% Through this bidirectional fusion, the MLLM-side embeddings absorb fine-grained textual details, while the original-text embeddings gain complementary coarse-grained multimodal semantics, resulting in mutually enriched and complementary item representations.

\begin{algorithm}[!t]
\caption{Training and inference process of DMESR}
\label{alg:train}
\raggedright
\textbf{Initialization}
\begin{algorithmic}[1]
    \State Initialize adapter networks ($\text{Adapter}_1$, $\text{Adapter}_2$), cross-attention modules, and SRS backbone $f_{\theta}$.
    \State Generate descriptions and encode both MLLM's and original text: $\{\mathbf{e}_{v_i}^t, \mathbf{e}_{v_i}^i, \mathbf{e}_{v_i}^h,\mathbf{e}_{v_i}^{T}\}_{v_i \in \mathcal{V}}$.
\end{algorithmic}

\textbf{Training Process}
\setcounter{algorithm}{2}
\begin{algorithmic}[1]
    \makeatletter
    \setcounter{ALG@line}{3}
    \While{Not converged}
        \For{each batch of users $\mathcal{B} \subseteq \mathcal{U}$}
            \State Feed three routes into $\text{Adapter}_1$ and compute $\mathcal{L}_{align}$ by Eq.~\eqref{eq:contrastive_loss} to get $\mathbf{e}_{v_i}^{coarse}$.
            \State Feed original text into $\text{Adapter}_2$ to get $\mathbf{e}_{v_i}^{fine}$.
            \State Perform bidirectional cross-attention on $\mathcal{E}_u^{coarse}$ and $\mathcal{E}_u^{fine}$ by Eq.~\eqref{eq:cross_attn} to obtain $\hat{\mathcal{E}}_u^{coarse}$ and $\hat{\mathcal{E}}_u^{fine}$.
            \State Feed $\hat{\mathcal{E}}_u^{coarse}$ and $\hat{\mathcal{E}}_u^{fine}$ into SRS backbone $f_{\theta}$ to get $\mathbf{u}^{coarse}$ and $\mathbf{u}^{fine}$.
            \State Calculate prediction probability by Eq.~\eqref{eq:predict} and compute $\mathcal{L}_{SRS}$ by Eq.~\eqref{eq:SRS_loss}.
            \State Calculate total loss: $\mathcal{L} = \mathcal{L}_{SRS} + \alpha \mathcal{L}_{align}$.
            \State Update parameters $\Theta$ and $\Phi$.
        \EndFor
    \EndWhile
\end{algorithmic}

\textbf{Inference Process}
\setcounter{algorithm}{11}
\begin{algorithmic}[1]
    \makeatletter
    \setcounter{ALG@line}{12}
    \State Load precomputed embeddings: $\{\mathbf{e}_{v_i}^t, \mathbf{e}_{v_i}^i, \mathbf{e}_{v_i}^h, \mathbf{e}_{v_i}^{T}\}_{v_i \in \mathcal{V}}$.
    \State Feed embeddings through adapters, cross-attention, and SRS backbone and Calculate prediction probability by Eq.~\eqref{eq:predict}.
\end{algorithmic}
\end{algorithm}
% \vspace{-8pt}

% -------------------------------
% \subsubsection{Collaborative Recommendation.}
% \vspace{2mm}
\subsubsection{\textbf{Collaborative Recommendation}}
To leverage the two complementary embeddings after bidirectional fusion, we adopt a deep sequential recommender to separately model user preferences from each view and train them collaboratively.

For the user $u$ with the two types of cross-fused sequence of item embeddings: $\hat{\mathcal{S}}_u^{coarse}=\{\hat{\mathbf{e}}_{v_1}^{coarse}, \hat{\mathbf{e}}_{v_2}^{coarse}, \ldots, \hat{\mathbf{e}}_{v_{n_u}}^{coarse}\}$ and $\hat{\mathcal{S}}_u^{fine}=\{\hat{\mathbf{e}}_{v_1}^{fine}, \hat{\mathbf{e}}_{v_2}^{fine}, \ldots, \hat{\mathbf{e}}_{v_{n_u}}^{fine}\}$.  
We feed these two embedding sequences into sequential recommendation model separately to capture their user preferences: $\mathbf{u}^{coarse}={f_{\theta}}(\hat{\mathcal{S}}_u^{coarse})$ and $\mathbf{u}^{fine}={f_{\theta}}(\hat{\mathcal{S}}_u^{fine})$, where $f_{\theta}(\cdot)$ denotes the backbone of SRS. During the prediction phase, we concatenate the user and item embeddings from both views to form the final representation for recommendation, and the probability score of recommending item $i$ to user $u$ is computed as:

\begin{equation}
\label{eq:predict}
P(v_{n_u+1}=v_i \mid v_{1:{n_u}})=[\hat{\mathbf{e}}_{v_i}^{coarse}:\hat{\mathbf{e}}_{v_i}^{fine}]^{\top}[\mathbf{u}^{coarse}:\mathbf{u}^{fine}]
\end{equation}

The sequential recommendation model is optimized with pairwise binary cross-entropy(BCE) loss:
\begin{equation}
\label{eq:SRS_loss}
\begin{aligned}
\mathcal{L}_{\text{SRS}}
= - \sum_{u \in \mathcal{U}} \sum_{k=1}^{n_u-1}
\Big[ 
& \log P(v_{k+1}=v_{k+1}^+ \mid v_{1:k}) \\
& + \log \big(1 - P(v_{k+1}=v_{k+1}^- \mid v_{1:k})\big)
\Big]
\end{aligned}
\end{equation}
where $v_{k+1}^+$ and $v_{k+1}^-$ are the ground-truth and paired negative item.

\subsection{Training and Inference}
\label{sec:method_train}

We present the training and inference procedures of DMESR in Algorithm~\ref{alg:train}. The framework is initialized by setting up the adapter networks, cross-attention modules, and SRS backbone (line 1). The MLLM generates three routes of descriptions with predefined prompts, then both the MLLM-generated descriptions and original text are encoded into embeddings for subsequent use (line 2).

\noindent\textbf{Training.}
During training, the three routes of embeddings are fed into $\text{Adapter}_1$ for dimensionality reduction, and the contrastive alignment loss $\mathcal{L}_{align}$ is computed to obtain the coarse-grained multimodal semantic embeddings (line 6). Meanwhile, the fine-grained semantic embeddings are derived through $\text{Adapter}_2$(line 7). Then, bidirectional cross-attention is performed to obtain the fused semantic embeddings (line 9), which are fed into the SRS backbone for prediction and loss computation (lines 10-11). The overall optimization objective is:
\begin{equation}
\arg\min_{\Theta,\Phi} \big( \mathcal{L}_{SRS} + \alpha \mathcal{L}_{align} \big)
\end{equation}
where $\Theta$ denotes the parameters of the SRS backbone and cross-attention modules, $\Phi = \{\mathbf{W}_1, \mathbf{W}_2, \mathbf{b}_1, \mathbf{b}_2\}$ represents the parameters of the adapters, and $\alpha$ is a hyperparameter that controls the strength of the alignment.

\noindent\textbf{Inference.}
At inference, all item embeddings are precomputed and loaded (line 13). They are sequentially passed through the adapters, cross-attention modules, and the SRS backbone to produce recommendation results (line 14). %The adapter and cached embeddings function as the \textit{\textbf{Embedding}} component of the model and introduce only slight computational overhead compared with traditional SRS models.

\section{Experiment}

\subsection{Experimental Settings}

% \textbf{Datasets.}  
% Three real-world datasets are adopted for evaluation, namely MovieLens, Yelp, and Amazon Games.  
% We follow the previous SRS works for preprocessing and data split~\cite{kang2018self,tang2018personalized}.  
% Each dataset is organized into user interaction sequences, and Items with insufficient interaction records are filtered out, as we do not consider the cold-start problem.
% The detailed statistics of the three datasets are summarized in Table~\ref{tab:dataset_statistics}.

\begin{table}[H]
\centering
\tabcolsep=0.1cm 
\caption{Statistics of the three benchmark datasets.}
% \small
% \renewcommand{\arraystretch}{1.1}
% \setlength{\tabcolsep}{3pt}
\resizebox{1\linewidth}{!}{
\begin{tabular}{lccccc}
\toprule
\textbf{Dataset} & \textbf{\#Users} & \textbf{\#Items} & \textbf{\#Inter.} & \textbf{Avg. Seq Len} & \textbf{Sparsity} \\
\midrule
MovieLens & 610 & 9,722 & 100,808 & 165.26 & 98.30\% \\
Yelp & 31,350 & 10,562 & 171,835 & 5.48 & 99.95\% \\
Amazon Games & 26,574 & 11,752 & 227,774 & 8.57 & 99.93\% \\
\bottomrule
\end{tabular}
}
\label{tab:dataset_statistics}
% \vspace{-6mm}
\end{table}

\subsubsection{\textbf{Datasets}}  
We evaluate our proposed DMESR framework on three widely used real-world benchmark datasets, namely MovieLens, Yelp, and Amazon Games. 
MovieLens is a movie rating dataset collected by the GroupLens research group, and we adopt the widely used \texttt{ml-latest-small} version\footnote{\url{https://grouplens.org/datasets/movielens/}}.
Yelp is a large-scale business review dataset, where we use the restaurant-related interactions from the Yelp Open Dataset\footnote{\url{https://www.yelp.com/dataset}}.
Amazon Games is a sub-category of the Amazon product review dataset, which contains users’ interactions with video game products on an e-commerce platform\footnote{\url{https://cseweb.ucsd.edu/~jmcauley/datasets.html\#amazon_reviews}}.

Following prior sequential recommendation studies~\cite{kang2018self,tang2018personalized}, each dataset is preprocessed and organized into user interaction sequences according to timestamps.
For preprocessing, we apply dataset-specific filtering strategies to obtain reliable user behavior sequences and avoid the cold-start issue. 
For the Amazon Games dataset, users with fewer than $5$ interactions and items with fewer than $3$ interaction records are removed. 
For the Yelp dataset, users with fewer than $4$ interactions and items with fewer than $3$ interactions are filtered out. 
For MovieLens, we do not apply additional filtering due to its relatively dense interactions and long sequential characteristics.
For training and evaluation, we adopt the leave-one-out protocol in sequential recommendation. 
Specifically, for each user sequence, the penultimate interaction is used as the validation instance, while the last interaction is reserved for testing, and all preceding interactions are used for training. 
The detailed statistics of the processed datasets are summarized in Table~\ref{tab:dataset_statistics}.

\begin{table*}[t]
\centering
\caption{Overall performance of competing baseline and our proposed DMESR under three backbone models across three datasets.
The best results are shown in \textbf{bold}, and the second-best are \underline{underlined}. ``\textbf{{\Large *}}'' indicates the statistically significant improvements (i.e., two-sided t-test with $p<0.05$) over the best baseline.}
%\scriptsize
% \renewcommand{\arraystretch}{1.05}
% \setlength{\tabcolsep}{10pt}
\resizebox{0.65\linewidth}{!}{
\begin{tabular}{llcccccc}
\toprule
%\multicolumn{2}{c}{\multirow{2}{*}{\textbf{Backbone}}} &
\multicolumn{1}{c}{\multirow{2}{*}{\textbf{Backbone}}} &
\multicolumn{1}{c}{\multirow{2}{*}{\textbf{Baseline}}} &
\multicolumn{2}{c}{\textbf{MovieLens}} &
\multicolumn{2}{c}{\textbf{Yelp}} &
\multicolumn{2}{c}{\textbf{Amazon Games}} \\
\cmidrule(lr){3-4} \cmidrule(lr){5-6} \cmidrule(lr){7-8}
 &  & H@10 & N@10 & H@10 & N@10 & H@10 & N@10 \\
\midrule

% ================= SASRec =====================
\multirow{5}{*}{\textbf{SASRec}} 
 & - None & 0.4951 & 0.3307 & 0.6346 & 0.4373 & 0.5574 & 0.3710 \\
 & - CPMM & \underline{0.8426} & \underline{0.6259} & \underline{0.8139} & \underline{0.5229} & 0.6387 & 0.4186 \\
 & - QARM & 0.8082 & 0.5883 & 0.7227 & 0.4888 & \underline{0.6716} & \textbf{0.4544} \\
 & - X-REFLECT & 0.7426 & 0.5034 & 0.8039 & 0.5016 & 0.6334 & 0.4023 \\
 & - \textbf{DMESR} & \textbf{0.8557*} & \textbf{0.6361*} & \textbf{0.8246*} & \textbf{0.5471*} & \textbf{0.6717} & \underline{0.4500} \\
\midrule

% ================= Bert4Rec =====================
\multirow{5}{*}{\textbf{Bert4Rec}} 
 & - None & 0.6738 & 0.4270 & 0.7084 & 0.4596 & 0.5545 & 0.3527 \\
 & - CPMM & \underline{0.7787} & \underline{0.5410} & \underline{0.7890} & \underline{0.4909} & 0.6045 & 0.3862 \\
 & - QARM & 0.7115 & 0.4529 & 0.7476 & 0.4895 & \underline{0.6556} & \underline{0.4457} \\
 & - X-REFLECT & 0.6803 & 0.4318 & 0.7821 & 0.4708 & 0.6147 & 0.3833 \\
 & - \textbf{DMESR} & \textbf{0.7967*} & \textbf{0.5517*} & \textbf{0.8102*} & \textbf{0.5241*} & \textbf{0.6725*} & \textbf{0.4507*} \\
\midrule

% ================= GRU4Rec =====================
\multirow{5}{*}{\textbf{GRU4Rec}} 
 & - None & 0.4951 & 0.3462 & 0.5259 & 0.3399 & 0.5081 & 0.3092 \\
 & - CPMM & \underline{0.5803} & \underline{0.3860} & 0.6750 & 0.4084 & 0.4622 & 0.2753 \\
 & - QARM & 0.5164 & 0.3593 & 0.5567 & 0.3650 & 0.6048 & \underline{0.3850} \\
 & - X-REFLECT & 0.5721 & 0.3550 & \underline{0.7784} & \underline{0.4710} & \underline{0.6170} & 0.3823 \\
 & - \textbf{DMESR} & \textbf{0.6180*} & \textbf{0.4409*} & \textbf{0.8042*} & \textbf{0.5076*} & \textbf{0.6402*} & \textbf{0.4164*} \\
\bottomrule
\end{tabular}
}
\label{tab:overall_comparison}
% \vspace{-3mm}
\end{table*}

\subsubsection{\textbf{Baselines}}  
To validate the generality and flexibility of the proposed framework, we integrate both our DMESR and the competing baselines with three widely used SRS backbone: SASRec~\cite{sun2019bert4rec}, BERT4Rec~\cite{kang2018self}, and GRU4Rec~\cite{hidasi2015session}.
In the experiments, we compare DMESR with three state-of-the-art MLLM-based item-side semantic enhancement methods:
\begin{itemize}[leftmargin=*]
    \item \textbf{X-REFLECT}~\cite{lyu2024x} prompts MLLMs to jointly analyze textual and visual content of items, focusing on the consistency and complementarity between modalities, and concatenates the generated multimodal descriptions to obtain enhanced item embeddings.  
    \item \textbf{CPMM}~\cite{mo2025one} obtains ID embeddings by summing the textual and visual representations of each item, and injects Gaussian noise for semantic purification, followed by contrastive learning, L2-shift loss, and co-occurrence-based enhancement to achieve more consistent item semantics.
    \item \textbf{QARM}~\cite{luo2024qarm} fine-tunes MLLM-enhanced embeddings via contrastive learning and further applies vector quantization (VQ) and residual quantization (RQ) to map continuous embeddings into discrete ID codes, enabling efficient ID-based recommendation.  
\end{itemize}
% , inlcuding X-REFLECT~\cite{lyu2024x}, CPMM~\cite{mo2025one}, and  QARM~\cite{luo2024qarm}, which all utilize MLLMs to enrich item representations from textual and visual modalities.  
% All baselines are implemented under the same experimental settings for fair comparison. 
%The more details about baselines are put into Apendix.
% \vspace{1mm}

% \noindent\textbf{Implementation Details.}
% All experiments are conducted on a HiSilicon Kunpeng-920 server equipped with one NVIDIA A100-PCIE-40GB GPU.
% The environment is configured with Python~3.10.18 and PyTorch~2.5.0+cu118. For the MovieLens dataset, we set the batch size to 128, while for the Yelp and Amazon Games datasets, the batch size is set to 1024. As for the MLLM, we choose Qwen2.5-VL-3B~\cite{bai2025qwen2} for this study.
% For the contrastive alignment strength $\alpha$, we search from $\{0.01, 0.05, 0.1, 0.5, 1\}$ to determine the optimal value.
% The implementation code is publicly available at:
% \url{https://anonymous.4open.science/r/DMESR-3141}.
% \vspace{1mm}

\subsubsection{\textbf{Implementation Details}}
All experiments are conducted on a HiSilicon Kunpeng-920 server equipped with one NVIDIA A100-PCIE-40GB GPU.
The software environment is configured with Python~3.10.18, PyTorch~2.5.0+cu118, and CUDA~12.2.
For the MovieLens dataset, the batch size is set to $128$, while for the Yelp and Amazon Games datasets, a larger batch size of $1024$ is adopted to improve training efficiency on sparse data.
Adam is used as the optimizer for all experiments, together with an early stopping strategy with a patience of $20$ epochs to prevent overfitting.
For multimodal semantic enhancement, we employ Qwen2.5-VL-3B~\cite{bai2025qwen2} as the MLLM backbone to generate enriched item descriptions.
The embeddings of the generated textual descriptions are obtained using the \texttt{text-ada-embedding-002} API.
For the proposed DMESR architecture, the bidirectional cross-attention module consists of a single cross-attention layer, and the embedding dimension is fixed to a hidden size of $128$ for all backbone SRS models.
For contrastive learning, the temperature parameter $\tau$ is set to $2$, while the contrastive alignment strength $\alpha$ is searched from $\{0.01, 0.05, 0.1, 0.5, 1\}$ to determine the optimal configuration for each dataset.
The implementation code is publicly available at:
\url{https://github.com/mingyao-huang/DMESR.git}.

\subsubsection{\textbf{Evaluation Metrics}}  
We evaluate all models under the standard \textit{Top-10} ranking setting.  
Specifically, two commonly used metrics are adopted: \textit{Hit Rate} (\textbf{H@10}) and \textit{Normalized Discounted Cumulative Gain} (\textbf{N@10}).  
Following~\cite{kang2018self}, for each user, we randomly sample 100 items that the user has not interacted with as negative candidates, combined with the ground-truth positive item to compute the evaluation metrics.  
%Both overall and tail-item performance are calculated to comprehensively verify the effectiveness of the proposed model.
% \vspace{-8pt}

\subsection{Overall Performance}

To validate the effectiveness and flexibility of {DMESR}, we report the overall performance of our method and all baselines under three sequential backbone models on three datasets, as shown in Table~\ref{tab:overall_comparison}.  
% In general, DMESR outperforms all competing methods across both evaluation metric, demonstrating its strong capability in enhancing sequential recommendation through the integration of coarse- and fine-grained semantics. 
% In the following, we present detailed analyses from two perspectives.
% obtained from MLLMs and original texts. 

\vspace{1mm}
\noindent\textbf{Overall Comparison.}  
%Generally, the proposed DMESR achieves the best performance under all datasets and backbone models on both metrics, verifying its superior effectiveness.  
Overall, DMESR achieves the best or highly competitive performance across datasets and backbone models on both metrics, demonstrating its strong effectiveness.
%Among the baselines, CPMM generally performs second-best, even though it also adopts contrastive learning.  
Among the baselines, CPMM generally achieves the second-best performance, as it employs contrastive learning together with several semantic alignment mechanisms to produce more consistent item representations, but the lack of fine-grained semantics may limit its further performance.
%This observation suggests that incorporating fine-grained textual semantics indeed enhances the representational quality of items in sequential recommendation systems.
By contrast, X-REFLECT performs relatively worse, as it directly concatenates MLLM-generated embeddings from different modalities without considering the inherent semantic gaps among them.  
For QARM, it achieves competitive results on the SASRec model, which can be attributed to its quantization of item embeddings into discrete codes, making it more aligned with ID-based recommendation, where SASRec excels.

\vspace{1mm}
\noindent\textbf{Flexibility.}  
% Table~\ref{tab:overall_comparison} shows that DMESR can achieve the largest performance gains across all three SRS backbone models, validating its strong adaptability and flexibility.  
Table~\ref{tab:overall_comparison} shows that DMESR achieves the best performance in 17 out of 18 settings, demonstrating superior adaptability across diverse recommendation scenarios.
Quantitatively, DMESR achieves an average improvement of \textbf{31.21\%} on H@10 and \textbf{35.68\%} on N@10 over the None baseline, substantially outperforming CPMM (17.35\% and 20.75\%), QARM (23.6\% and 19.4\%), and X-REFLECT (21.95\% and 16.93\%).
In comparison, other methods tend to show improvement depending on the specific type of SRS backbone---CPMM performs well with SASRec and BERT4Rec, QARM achieves notable gains on SASRec, while X-REFLECT shows relative advantages under GRU4Rec.  
%\lqd{Why? Explain the reason why our model owns better adaptability and flexibility.}

\subsection{Ablation Study}

\begin{table}[t]
\tabcolsep=0.1cm 
\centering
\caption{Ablation study of DMESR under three backbone models on the MovieLens dataset. 
The \textbf{bold} refers to the highest scores, while the \underline{underlined} indicate the second-best performance.}
%\scriptsize
% \small
% \renewcommand{\arraystretch}{1.1}
% \setlength{\tabcolsep}{5pt}
\resizebox{1\linewidth}{!}{
\begin{tabular}{l|cc|cc|cc}
\toprule
\multirow{2}{*}{\textbf{Model}} &
\multicolumn{2}{c|}{\textbf{SASRec}} &
\multicolumn{2}{c|}{\textbf{Bert4Rec}} &
\multicolumn{2}{c}{\textbf{GRU4Rec}} \\
\cmidrule(lr){2-3} \cmidrule(lr){4-5} \cmidrule(lr){6-7}
 & \textbf{H@10} & \textbf{N@10} & \textbf{H@10} & \textbf{N@10} & \textbf{H@10} & \textbf{N@10} \\
\midrule
\textbf{DMESR} & \textbf{0.8557} & \underline{0.6361} & \textbf{0.7967} & \textbf{0.5517} & \underline{0.6180} & \textbf{0.4409} \\
-\textit{w/o} CL        & 0.8180 & 0.6021 & 0.7066 & 0.4689 & {0.5705} & 0.4031 \\
-\textit{w/o} CA     & \underline{0.8457} & \textbf{0.6391} & \underline{0.7754} & \underline{0.5473} & 0.6016 & 0.4031 \\
-\textit{w/o} Ori-view       & 0.8230 & 0.6109 & 0.7607 & 0.5175 & 0.5754 & 0.3960 \\
-\textit{w/o} TWP     & 0.8393 & 0.6253 & 0.7164 & 0.4911 & \textbf{0.6328} & \underline{0.4243} \\
\bottomrule
\end{tabular}
}
\label{tab:ablation_DMESR}
% \vspace{-1mm}
\end{table}

% The ablation results of DMESR are reported in Table~\ref{tab:ablation_DMESR}.  
% First, we remove the contrastive learning module, denoted as \textit{w/o CL}, and directly concatenate the three embeddings.  
% This modification leads to a significant decline in performance across all backbones, indicating that the absence of contrastive alignment results in inconsistent semantic spaces among modalities, thereby degrading item representation quality.  
% Next, \textit{w/o CA} means remove the cross-attention fusion module, it also causes noticeable drops in most metrics, demonstrating that the bidirectional semantic interaction between the two semantic views plays an essential role in representation enhancement.  
% The results of these two variants together validate the necessity of the key components designed in DMESR.

% To further investigate the contributions of the fine-grained semantics and the three-way prompting framework, we remove the original semantic view, denoted as \textit{w/o Ori-view}, and remove the image and hybrid routes to eliminate the three-way prompting framework, denoted as \textit{w/o TWP}.  
% Both variants show performance degradation across most SRS backbone, confirming that incorporating fine-grained semantics and employing the three-way prompting framework are crucial for capturing complementary and multimodal item information.

The ablation results of DMESR are reported in Table~\ref{tab:ablation_DMESR}.  
First, we remove the contrastive learning module, denoted as \textit{w/o CL}, and directly concatenate the three embeddings.  
This modification leads to a significant decline in performance across all backbones, indicating that the absence of contrastive alignment results in inconsistent semantic spaces among modalities, thereby degrading item representation quality.  
Next, \textit{w/o CA} removes the cross-attention fusion module, which causes noticeable drops in most metrics, demonstrating that the bidirectional semantic interaction between the two semantic views plays an essential role in representation enhancement.
Although \textit{w/o CA} shows marginal improvements on SASRec N@10 (0.6361 $\rightarrow$ 0.6391), this is attributed to reduced model complexity that alleviates minor overfitting on this specific metric-backbone combination, while the overall performance degradation across other settings validates the necessity of cross-attention fusion.
The results of these two variants together confirm the importance of the key components designed in DMESR.

To further investigate the contributions of the fine-grained semantics and the three-way prompting framework, we remove the original semantic view, denoted as \textit{w/o Ori-view}, and remove the image and hybrid routes to eliminate the three-way prompting framework, denoted as \textit{w/o TWP}.  
Both variants generally show performance degradation across most backbone-metric pairs, confirming that incorporating fine-grained semantics and employing the three-way prompting framework are crucial for capturing complementary and multimodal item information.
Notably, \textit{w/o TWP} achieves higher H@10 on %GRU4Rec (0.6180 $\rightarrow$ 0.6328), which can be explained by the fact that GRU4Rec's sequential modeling capacity is hindered by redundancy with additional prompting routes.
Notably, \textit{w/o TWP} achieves higher H@10 on GRU4Rec (0.6180 $\rightarrow$ 0.6328), as the simplified single-route architecture better aligns with GRU4Rec's relatively simple sequential modeling capacity, reducing the overhead from multi-route fusion by contrastive learning.
% , and the simplified single-route approach reduces this redundancy for the H@10 metric.

\vspace{-1mm}

\subsection{Long-tail Item Analysis}

\begin{figure}[t]
    \centering

    % ---------- 第一张（MovieLens） ----------
    \begin{subfigure}[t]{\linewidth}
        %\centering
        \hspace{-0.02\columnwidth}
        \includegraphics[width=1.04\linewidth]{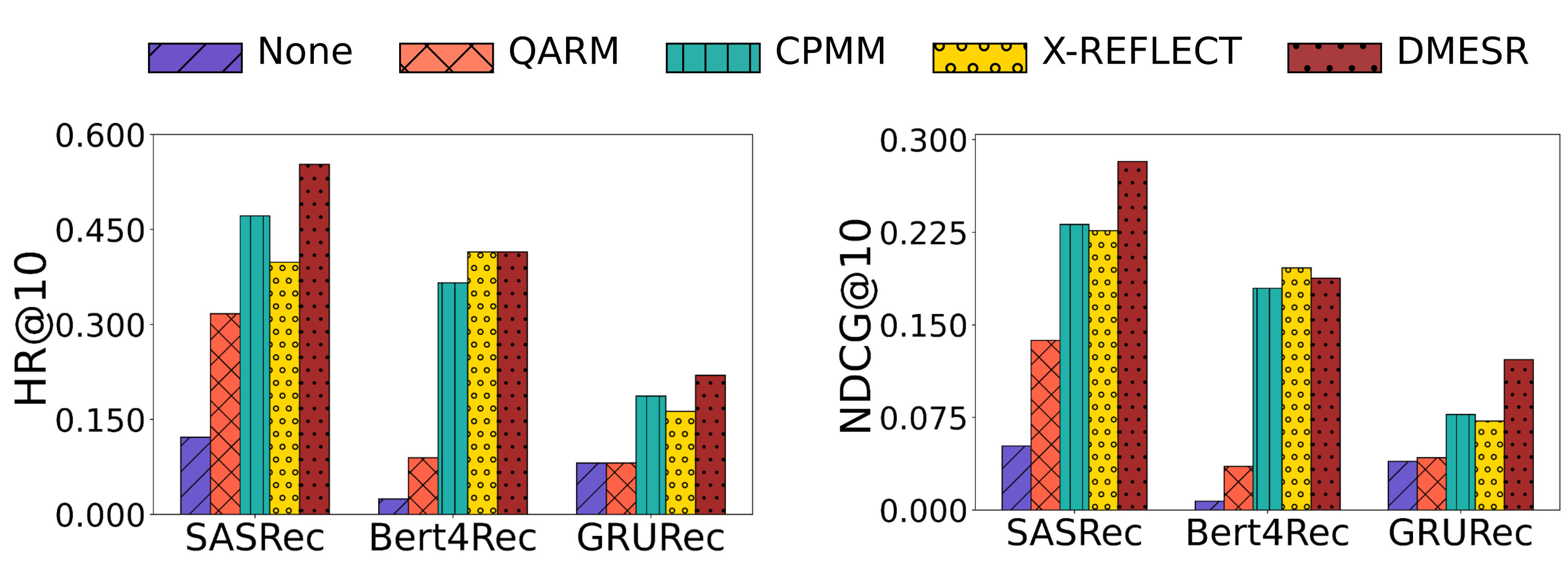}
        \caption{MovieLens}
    \end{subfigure}

    % \vspace{3mm}

    % ---------- 第二张（Yelp） ----------
    \begin{subfigure}[t]{\linewidth}
        % \centering
        \hspace{-0.02\columnwidth}
        \includegraphics[width=1.04\linewidth]{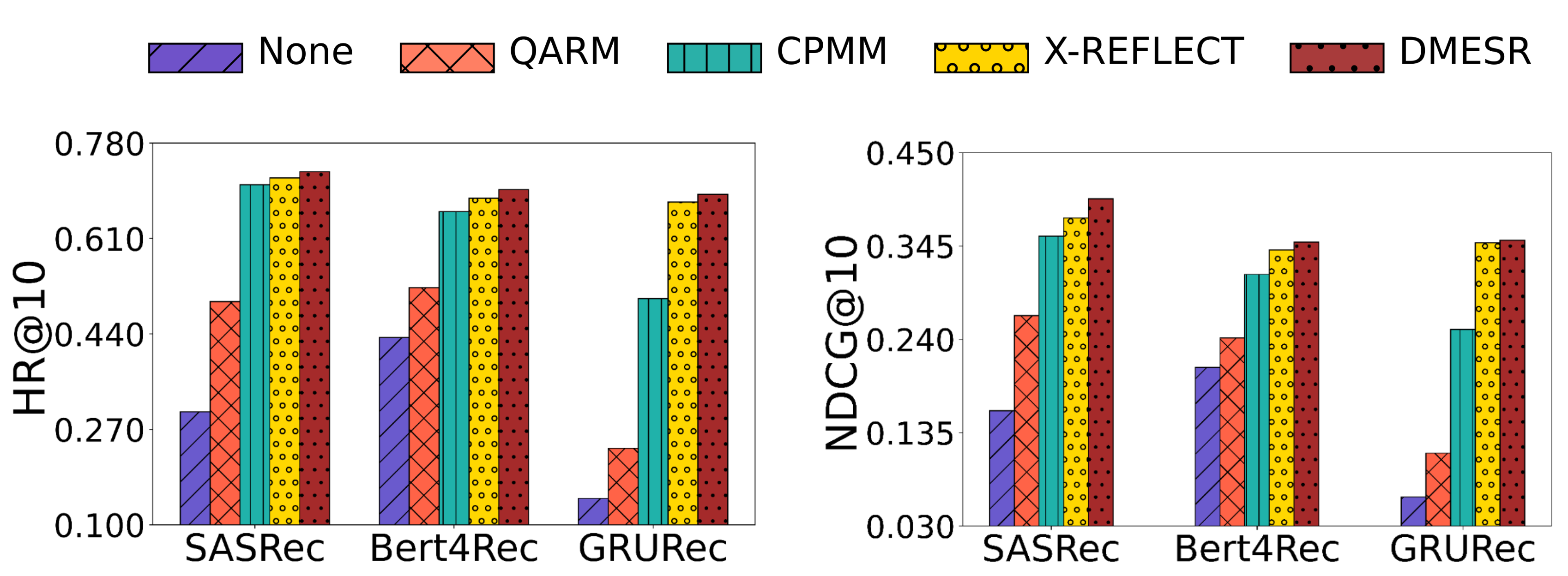}
        \caption{Yelp}
    \end{subfigure}

    \caption{Experimental results of long-tail items on MovieLens and Yelp datasets.}
    \label{fig:long_tail_item_result}
\vspace{-2mm}
\end{figure}

To further evaluate the performance of DMESR on long-tail items, we conduct experiments on the MovieLens and Yelp datasets under three SRS backbone.  
According to the Pareto principle~\cite{box1986analysis}, we define the last 80\% of items as long-tail items, and the results of them are presented in Figure~\ref{fig:long_tail_item_result}.  
As illustrated, DMESR generally achieves the best performance across both metrics and SRS backbone, demonstrating that the proposed dual-view enhancing framework can effectively improve recommendation accuracy not only in overall performance but also on long-tail items.
Among the baselines, X-REFLECT shows the second-best performance, indicating that its concatenation-based multimodal enhancement can effectively improve long-tail item representation.
In contrast, QARM performs worst among the three baselines, as its ID-based quantized embeddings struggle to effectively enrich the item representations, limiting its ability to enhance recommendations for long-tail items.
\vspace{-2mm}

\subsection{Hyper-parameter Analysis}

\begin{figure}[t]
    \makebox[\columnwidth][l]{%
        \hspace*{-0.03\columnwidth}
        \includegraphics[width=1.04\columnwidth]{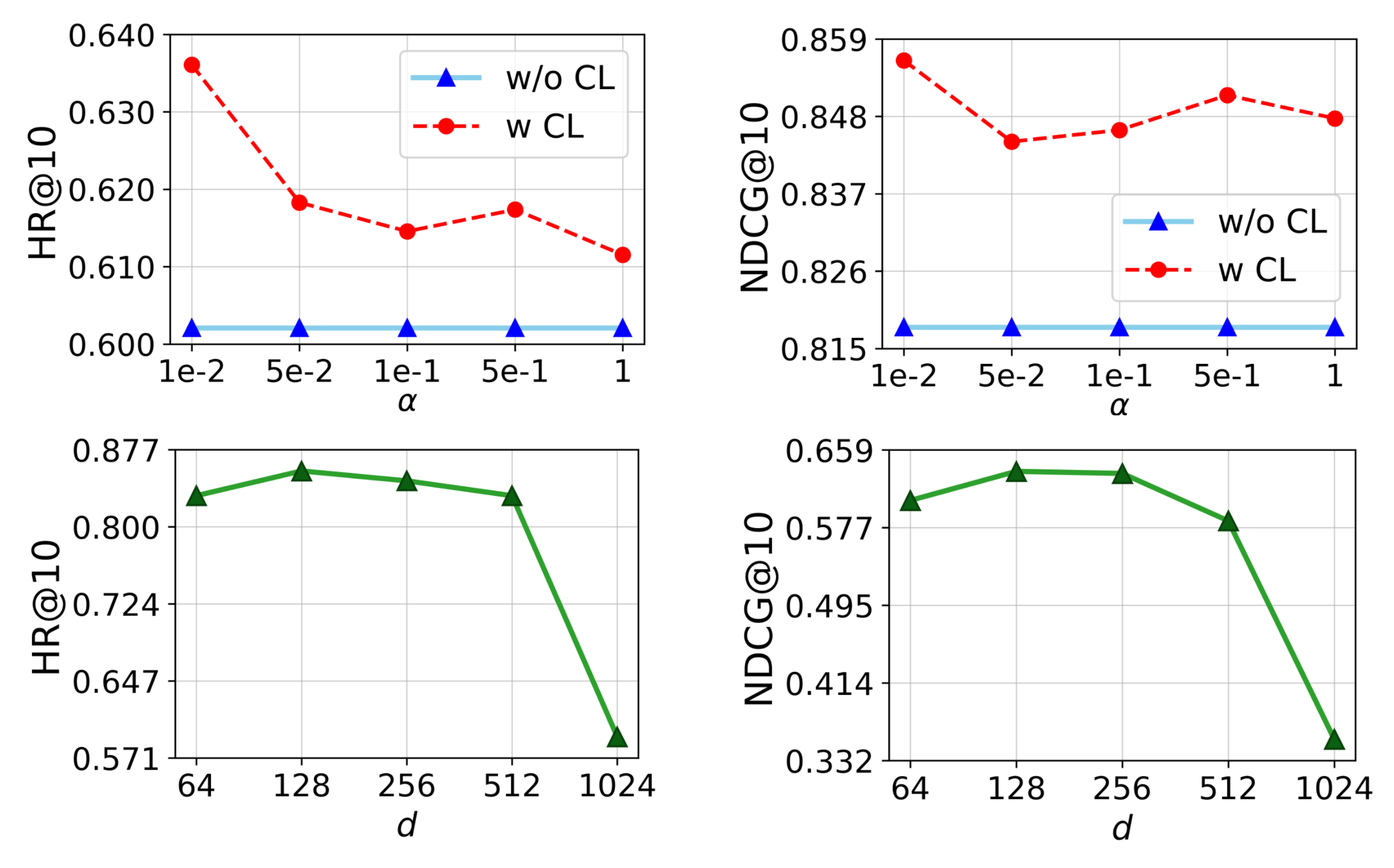}
    }
    \caption{The hyper-parameter experiments of the weight of contrastive learning loss $\alpha$ and the size of the embedding dimension $d$ on the MovieLens dataset with the SASRec model.}
    \label{fig:hyper_parameter_results}
\vspace{-2mm}
\end{figure}

To investigate the impact of key hyper-parameters in DMESR, we conduct sensitivity experiments on the contrastive learning weight $\alpha$ and the size of embedding dimension $d$. 
As shown in Figure~\ref{fig:hyper_parameter_results}, the contrastive learning weight $\alpha$ varies from 0.01 to 1, the overall recommendation accuracy gradually decreases. 
This is because large $\alpha$ overemphasis on the contrastive loss, which in turn affects the convergence of the prediction loss. Nevertheless, all variants still outperform the model without contrastive learning, validating the effectiveness of the contrastive alignment module. 
As for the embedding dimension $d$, the best is 128. The reason is larger dimensionality can capture richer item information, but overly high $d$ leads to insufficient training, degrading recommendation accuracy. 
% \vspace{-2mm}

\subsection{MLLMs series Analysis}

\begin{figure}[!t]
    \makebox[\columnwidth][l]{%
        \hspace{-0.02\columnwidth}
        \includegraphics[width=1.04\columnwidth]{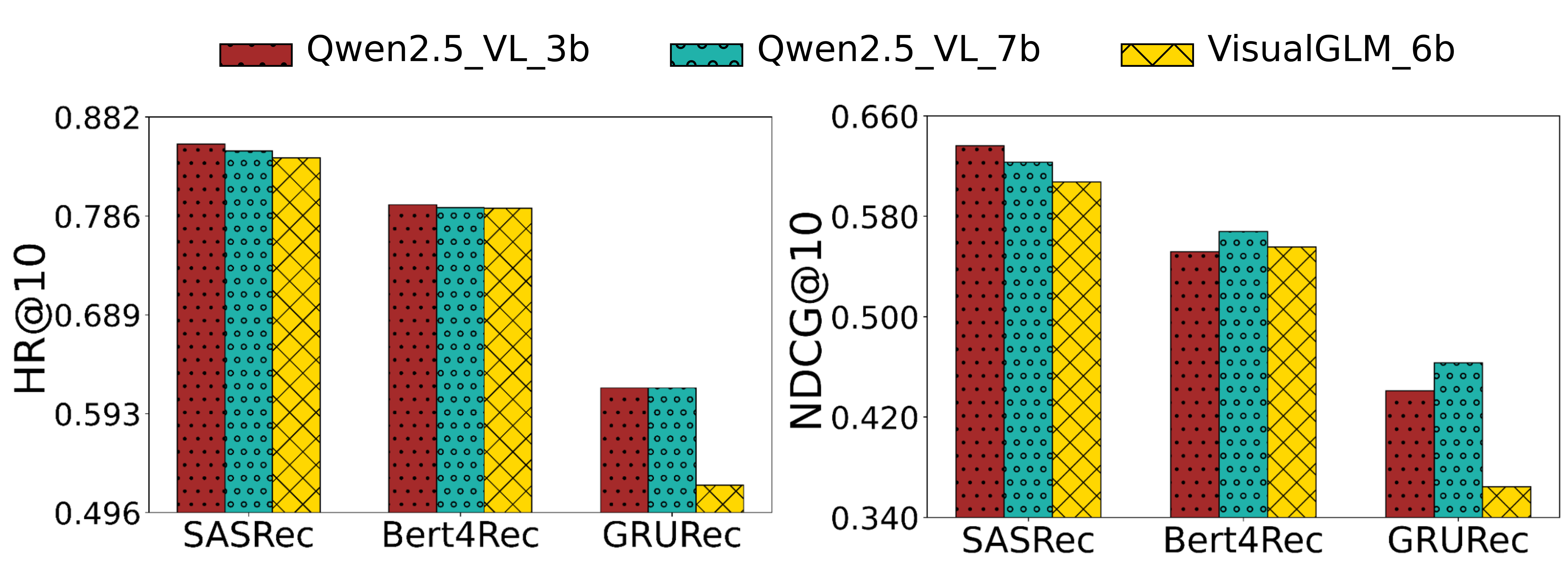}
    }
    \caption{Experimental results of three MLLMs on MovieLens.}
    \label{fig:LLM_Series_results}
% \vspace{-2mm}
\end{figure}

To investigate the adaptability of the proposed DMESR framework to different MLLMs, we conduct experiments from two perspectives: model size and model type, as shown in Figure~\ref{fig:LLM_Series_results}. 
For MLLMs of different sizes, we test Qwen2.5-VL-3B and Qwen2.5-VL-7B. The results indicate that DMESR achieves comparable performance across both models, with the 7B version showing slightly better results on GRU4Rec due to its stronger ability to generate richer item descriptions. 
For MLLMs of different type, we further introduce VisualGLM-6B for comparison. The performance remains similar overall, verifying the robustness of DMESR across different MLLM architectures, while a slight decline is observed for VisualGLM-6B on GRU4Rec, suggesting its reasoning and understanding abilities are relatively weaker than those of Qwen2.5-VL.

% \begin{figure}[!t]
%     % \centering
%     \hspace{-0.02\columnwidth}
%     \includegraphics[width=1.04\linewidth]{figures/LLM_Series.png}
%     \caption{Experimental results of three MLLMs on the MovieLens dataset.}
%     \label{fig:LLM_Series_results}
%     %\vspace{-6mm}
% \end{figure}

% \vspace{-2mm}
\section{Related Works}

% \textbf{Sequentail Recommendation.}
\subsection{\textbf{Sequentail Recommendation}}
Sequential recommendation aims to model users’ dynamic preferences from their interaction sequences to predict the next  likely item~\cite{liu2023linrec,liu2024sequential}.
Early methods such as GRU4Rec~\cite{hidasi2015session} and Caser~\cite{tang2018personalized} utilize RNNs and CNNs to capture temporal dependencies in user behaviors.
With the emergence of self-attention, Transformer-based models like SASRec~\cite{kang2018self} and BERT4Rec~\cite{sun2019bert4rec} further improved sequence modeling performance. More recently, GNN-based methods like TiDA-GCN~\cite{guo2022time} and REDREC~\cite{yao2023redrec} have been proposed to capture complex user–item transition patterns. Unlike these structure-oriented approaches, our DMESR enhances sequential recommendation from the item side by enriching item representations with multimodal semantics and can be seamlessly integrated into various backbone architectures.
\vspace{1mm}

% \noindent\textbf{MLLMs for Recommendation.}
\subsection{\textbf{MLLMs for Recommendation}}
With the rapid progress of MLLMs, leveraging their strong semantic understanding to enhance recommendation systems has become an emerging research focus.
One line of research directly employs MLLMs as recommenders. For example, RecGPT4V~\cite{liu2024rec} uses MLLMs to summarize image and rank candidate items through natural-language reasoning; MLLM-MSR~\cite{ye2025harnessing} models user preferences via an RNN-like interaction structure; and HistLLM~\cite{zhang2025histllm} encodes multimodal interaction histories into a single representation token for preference inference.
Although these methods demonstrate strong reasoning ability, they incur high computational costs during inference.
Another line of work leverages MLLMs as auxiliary tools to enhance traditional recommendation models during training. Many focus on item-side semantic enhancement, where MLLMs enrich item representations from textual and visual modalities. Representative works such as X-REFLECT~\cite{lyu2024x}, MMRec~\cite{tian2024mmrec}, CPMM~\cite{mo2025one}, and QARM~\cite{luo2024qarm} exploit multimodal comprehension to refine item embeddings and improve downstream recommendation performance. Our DMESR also follows this item-side enhancement paradigm, but unlike previous works, we integrate two complementary semantic views to achieve coarse- and fine-grained semantic fusion for item representation.
% \vspace{-2mm}
\section{Conclusion}

In this paper, we propose a dual-view MLLM-based enhancing framework for sequential recommendation (DMESR). 
We first apply a contrastive learning module to align the cross-modal embeddings produced by the three-way prompting framework, thus alleviating the semantic inconsistency problem.
Further, we introduce a bidirectional cross-attention module to integrate the coarse-grained semantics with the fine-grained semantics obtained from the original texts, thereby mitigating the issue of detailed semantic loss.
Through comprehensive experiments, we validate the effectiveness and flexibility of our DMESR.

\bibliographystyle{ACM-Reference-Format}
\bibliography{acmart}

\end{sloppypar}
\end{document}